\documentclass[11pt]{article}
\usepackage{amsmath}
\usepackage{amsthm}
\usepackage{titlesec}
\usepackage{indentfirst}

\usepackage{amsmath}
\usepackage{amsthm}
\usepackage{emlines}
\usepackage{emlines2}

\theoremstyle{definition}\newtheorem{Df}{Definition}
\theoremstyle{plain}\newtheorem{Th}{Theorem}
\theoremstyle{definition}\newtheorem{Rm}{Remark}
\theoremstyle{definition}\newtheorem{Emp}{Example}
\theoremstyle{plain}\newtheorem{Pp}[Th]{Proposition}
\theoremstyle{plain}\newtheorem{Co}[Th]{Corollary}
\theoremstyle{plain}\newtheorem{Lm}[Th]{Lemma}

\usepackage{amsthm}
\usepackage{amsmath,amssymb}
\begin{document}
\title{{\bf Hierarchy and equivalence of multi-letter quantum finite automata\thanks{This
research is supported  by the National Natural Science Foundation
(Nos. 60573006, 60873055), the Research Foundation for the
Doctorial Program of Higher School of Ministry of Education (No.
20050558015),  Program for New Century Excellent Talents in
University (NCET) of China, and the Natural Science and
Engineering Research Council of Canada Grand \#OGP0041630.}}}
\author{Daowen Qiu$^{a,c,}$\thanks{
{\it E-mail address:} issqdw@mail.sysu.edu.cn (D. Qiu).} ,\hskip
3mm Sheng Yu$^b $\thanks{
{\it E-mail address:} syu@csd.uwo.ca (S. Yu).} \\
\small{{\it $^a$Department of
Computer Science, Zhongshan University, Guangzhou 510275, China}}\\
\small{{\it $^b$Department of
Computer Science, The  University of Western Ontario, }}\\
\small{{\it  London, Ontario, N6A 5B7, Canada}}\\
\small{{\it $^c$SQIG--Instituto de Telecomunica\c{c}\~{o}es, IST,
TULisbon, }}\\
\small{{\it  Av. Rovisco Pais 1049-001, Lisbon, Portugal}}\\
}
\date{ }
\maketitle

 \vskip 2mm \noindent {\bf Abstract}

Multi-letter {\it  quantum finite automata} (QFAs) were a new
one-way QFA model proposed recently by Belovs, Rosmanis, and
Smotrovs (LNCS, Vol. 4588, Springer, Berlin, 2007, pp. 60-71), and
they showed that multi-letter QFAs can accept with no error some
regular languages ($(a+b)^{*}b$) that are unacceptable by the
one-way QFAs. In this paper, we continue to study multi-letter
QFAs. We mainly focus on two issues: (1) we show that
$(k+1)$-letter QFAs are computationally more powerful than
$k$-letter QFAs, that is, $(k+1)$-letter QFAs can accept some
regular languages that are unacceptable by any $k$-letter QFA. A
comparison with the one-way QFAs is made by some examples; (2) we
prove that  a $k_{1}$-letter QFA ${\cal A}_1$ and another
$k_{2}$-letter QFA ${\cal A}_2$ are equivalent if and only if they
are $(n_{1}+n_{2})^{4}+k-1$-equivalent,  and the time complexity
of determining the equivalence of two multi-letter QFAs using this
method is $O(n^{12}+k^{2}n^{4}+kn^{8})$, where $n_{1}$ and $n_{2}$
are the numbers of states of ${\cal A}_{1}$ and ${\cal A}_{2}$,
respectively, and $k=\max(k_{1},k_{2})$.  Some other issues are
addressed for further consideration.


\vskip 2mm

\section*{1. Introduction}
Quantum computing is an intriguing and promising research field,
which touches on computer science, quantum physics, and
mathematics  \cite{Gru99,Hir04,CP02,CDS00}. To a certain extent,
quantum computing was motivated by the exponential speed-up of
Shor's quantum algorithm for factoring integers in polynomial time
\cite{Sho} and
 Grover's algorithm of searching in database of size $n$
with only $O(\sqrt{n})$ accesses \cite{Gro96}.

Quantum computers---the physical devices complying with the rules
of quantum mechanics were first considered by Benioff
\cite{Ben80}, and then suggested by Feynman \cite{Fey82}. By
elaborating and formalizing Benioff and Feynman's idea, in 1985,
Deutsch \cite{Deu85} re-examined the Church-Turing Principle and
defined {\it quantum Turing machines} (QTMs). Subsequently,
Deutsch \cite{Deu89} considered quantum network models. In 1993,
Yao \cite{Yao93} demonstrated the equivalence between QTMs and
quantum circuits. Quantum computation from the viewpoint of
complexity theory was first studied systematically by Bernstein
and Vazirani \cite{BV97}.

Another kind of simpler models of quantum computation is {\it
quantum finite automata} (QFAs), which can be thought of as
theoretical models of quantum computers with finite memory. This
kind of computing machines was first studied  by Moore and
Crutchfield \cite{MC00}, as well as by Kondacs and Watrous
\cite{KW97} independently. Then it was dealt with in depth by
Ambainis and Freivalds \cite{AF98}, Brodsky and Pippenger
\cite{BP02}, and the other authors (for example, see the
references in \cite{Gru99,QL08}). The study of QFAs is mainly
divided into two ways: one is {\it one-way quantum finite
automata} (1QFAs) whose tape heads only move one cell to right at
each computation step (1QFAs have been extensively studied
\cite{BMP03}), and the other is {\it two-way quantum finite
automata} (2QFAs), in which the tape heads are allowed to move
towards right or left, or to be stationary \cite{KW97}. (Notably,
Amano and Iwama \cite{AI99} dealt with a decidability problem
concerning an intermediate form called 1.5QFAs, whose tape heads
are allowed to move right or to be stationary; Hirvensalo
\cite{Hir07} investigated a decidability problem related to
one-way QFAs.) Furthermore, by considering the number of times of
the measurement in a computation, 1QFAs have two different forms:
{\it measure-once} 1QFAs (MO-1QFAs) proposed by Moore and
Crutchfield \cite{MC00}, and, {\it measure-many} 1QFAs (MM-1QFAs)
studied first by Kondacs and Watrous \cite{KW97}.

MM-1QFAs are strictly more powerful than MO-1QFAs
\cite{AF98,BMP03} (Indeed, $a^{*}b^{*}$ can be accepted by
MM-1QFAs with bounded error but not by any MO-1QFA with bounded
error). Due to the unitarity of quantum physics and finite memory
of finite automata, both MO-1QFAs and MM-1QFAs can only accept
proper subclasses of regular languages with bounded error (e.g.,
\cite{KW97,AF98,BP02,BMP03}). Indeed, it was shown that the
regular language $(a+b)^{*}b$ cannot be accepted by any MM-1QFA
with bounded error \cite{KW97}.

Recently, Belovs, Rosmanis, and Smotrovs \cite{BRS07} proposed a
new one-way QFA model, namely, multi-letter QFAs, that can be
thought of as a quantum counterpart of more restricted classical
one-way multi-head finite automata (see, for example,
\cite{Hro83}). Roughly speaking,  a $k$-letter QFA is not limited
to seeing only one, the just-incoming input letter, but can see
several earlier received letters as well. That is, the quantum
state transition which the automaton performs at each step depends
on the last $k$ letters received. For the other computing
principle, it is similar to the usual MO-1QFAs as described above.
Indeed, when $k=1$, it reduces to an MO-1QFA. Any  given
$k$-letter QFA can be simulated by some $k+1$-letter QFA. However,
we will prove that the contrary does not hold.  Belovs et al.
\cite{BRS07} have already showed that $(a+b)^{*}b$ can be accepted
by a 2-letter QFA but, as proved in \cite{KW97}, it cannot be
accepted by any MM-1QFA with bounded error. By ${\cal L}(QFA_{k})$
we denote the class of languages accepted with bounded error by
$k$-letter QFAs. In this paper, we will prove that ${\cal
L}(QFA_{k})\subset {\cal L}(QFA_{k+1})$ for $k=1,2,...$, where the
inclusion $\subset$ is proper. Therefore, $(k+1)$-letter QFAs are
computationally more powerful than $k$-letter QFAs.

As we know, determining the equivalence for computing models is a
very important issue in the theory of classical computation (see,
e.g., \cite{Paz71,Sal85,RS97,CK89,HK90,HHK06}). Concerning the
problem of determining the equivalence for QFAs, there exists some
work \cite{BP02} that deals with the simplest case---MO-1QFAs. For
quantum sequential machines (QSMs), Qiu \cite{Qiu02} gave a
negative outcome for determining the equivalence of QSMs, and then
Li and Qiu \cite{LQ06} further gave a method for determining
whether or not any two given QSMs are equivalent. This method
applies to determining the equivalence between any two MO-1QFAs
and also is different from the previous ones. For the equivalence
problem of MM-1QFAs, inspired by the work of \cite {Tze92} and
\cite {BMP03}, Li and Qiu \cite{LQ08} presented a polynomial-time
algorithm for determining whether or not any two given MM-1QFAs
are equivalent.

In this paper,  we will give a polynomial-time algorithm for
determining whether or not any two given $k_{1}$-letter QFA ${\cal
A}_{1}$ and $k_{2}$-letter QFA  ${\cal A}_{2}$ for accepting unary
languages are equivalent. More specifically, we prove that two
multi-letter QFAs ${\cal A}_1$ and ${\cal A}_2$, are equivalent if
and only if they are $(n_{1}+n_{2})^{4}+k-1$-equivalent,  where
$n_{1}$ and $n_{2}$ are the numbers of states of ${\cal A}_{1}$
and ${\cal A}_{2}$, respectively, $k=\max(k_{1},k_{2})$, and two
multi-letter QFAs over the same input alphabet $\Sigma$ are
$n$-equivalent if and only if the accepting probabilities of
${\cal A}_{1}$ and ${\cal A}_{2}$ are equal for the input strings
of length not more than $n$. This method, generalized
appropriately, may apply to dealing with more general cases.

The remainder of the paper is organized as follows. In Section 2,
we recall the definition of multi-letter QFAs and other related
definitions, and some related results are reviewed. In Section 3,
we  prove that ${\cal L}(QFA_{k})\subset {\cal L}(QFA_{k+1})$ for
$k=1,2,...$, where the inclusion $\subset$ is proper. More
precisely, we show that, for $k\geq 2$, regular language
$(a_{1}+a_{2}+\ldots +a_{k})^{*}a_{1}a_{2}\cdots a_{k-1}$ cannot
be accepted with bounded error by $(k-1)$-letter QFAs but can be
exactly accepted by some $k$-letter QFAs. In addition, we present
a number of examples to show the relation between multi-letter
QFAs and the usual one-way QFAs.

In Section 4, we concentrate on the equivalence issue. After
proving some useful lemmas, we prove that a $k_{1}$-letter QFA
${\cal A}_1$ and another $k_{2}$-letter QFA ${\cal A}_2$ for
accepting unary languages are equivalent if and only if they are
$(n_{1}+n_{2})^{4}+k-1$-equivalent, and the time complexity of
determining the equivalence of two multi-letter DFAs using this
method is $O(n^{12}+k^{2}n^{4}+kn^{8})$,  where $n=n_{1}+n_{2}$,
$n_{1}$ and $n_{2}$ are the numbers of states of ${\cal A}_{1}$
and ${\cal A}_{2}$, respectively, and $k=\max(k_{1},k_{2})$.
Finally, in Section 5  we address some related issues for further
consideration.

In general, symbols will be explained when they first appear.

\section*{2. Preliminaries}

In this section, we briefly review some definitions and related
properties that will be used in the sequel. For the details, we
refer to \cite{BRS07}.

First we recall $k$-letter deterministic finite automata
($k$-letter DFAs).

\begin{Df}[{\cite{BRS07}}]
 A $k$-letter deterministic finite automaton ($k$-letter DFA) is defined by a
 quintuple $(Q,Q_{acc},q_{0},\Sigma,\gamma)$, where $Q$ is a
 finite set of states, $Q_{acc}\subseteq Q$ is the set of
 accepting states, $q_{0}\in Q$ is the initial state, $\Sigma$ is
 a finite input alphabet, and $\gamma$ is a transition
 function that maps $Q\times T^{k}$ to $Q$, where
 $T=\{\Lambda\} \bigcup \Sigma$ and letter $\Lambda\notin \Sigma$
 denotes the  blank symbol (like a blank symbol in Turing machines
 \cite{Sal85}), and $T^{k}\subset T^{*}$ consists of all strings of length
 $k$.
\end{Df}

We describe the computing process of a $k$-letter DFA on an input
string $x$ in $\Sigma^{*}$, where
$x=\sigma_{1}\sigma_{2}\cdots\sigma_{n}$, and $\Sigma^{*}$ denotes
the set of all strings over $\Sigma$. The $k$-letter DFA has a
tape which contains the letter $\Lambda$ in its first $k-1$
position followed by the input string $x$. The automaton starts in
the initial state $q_{0}$ and has $k$ reading heads which
initially are on the first $k$ positions of the tape (clearly, the
$k$th head reads $\sigma_{1}$ and the other heads read $\Lambda$).
Then the automaton transfers to a new state as current state and
all heads move right a position in parallel. Now the $(k-1)$th and
$k$th heads point to $\sigma_{1}$ and $\sigma_{2}$, respectively,
and the others, if any, to $\Lambda$. Subsequently, the automaton
transfers to a new state and all heads move to the right. This
process does not stop until the $k$th head has read the last
letter $\sigma_{n}$. The input string $x$ is accepted if and only
if the automaton enters an accepting state after its $k$th head
reading the last letter $\sigma_{n}$.

Clearly, $k$-letter DFAs are not more powerful than DFAs. The
family of languages accepted by $k$-letter DFAs, for $k\geq 1$, is
exactly the family of regular languages.

For the sake of readability, we briefly recall the definitions of
MO-1QFAs and MM-1QFAs in the following.

An MO-1QFA is defined as a quintuple
$A=(Q,Q_{acc},|\psi_{0}\rangle,\Sigma,\{U(\sigma)\}_{\sigma\in\Sigma})$,
where $Q$ is a set of finite states, $Q_{acc}\subseteq Q$ is the
set of accepting states, $|\psi_{0}\rangle$ is the initial state
that is a superposition of the states in $Q$, $\Sigma$ is a finite
input alphabet, and $U(\sigma)$ is a unitary matrix for each
$\sigma\in\Sigma$.

As usual, we identify $Q$ with an orthonormal base of a complex
Euclidean space and every state $q\in Q$ is identified with a
basis vector, denoted by Dirac symbol $|q\rangle$ (a column
vector), and $\langle q|$ is the conjugate transpose of
$|q\rangle$. We describe the computing process for any given input
string $x=\sigma_{1}\sigma_{2}\cdots\sigma_{m}\in\Sigma^{*}$. At
the beginning the machine $A$ is in the initial state
$|\psi_{0}\rangle$, and upon reading $\sigma_{1}$. The
transformation $U(\sigma_{1})$ acts on $|\psi_{0}\rangle$. After
that, $U(\sigma_{1})|\psi_{0}\rangle$ becomes the current state
and the machine reads $\sigma_{2}$. The process continues until
the machine has read $\sigma_{m}$ ending in the state
$|\psi_{x}\rangle=U(\sigma_{m})U(\sigma_{m-1})\cdots
U(\sigma_{1})|\psi_{0}\rangle$. Finally, a measurement is
performed on $|\psi_{x}\rangle$ and the accepting probability
$p_{a}(x)$ is equal to
\[
p_{a}(x)=\langle\psi_{x}|P_{a}|\psi_{x}\rangle=\|P_{a}|\psi_{x}\rangle\|^{2}
\]
where $P_{a}=\sum_{q\in Q_{acc}}|q\rangle\langle q|$ is the
projection onto the subspace spanned by $\{|q\rangle: q_{i}\in
Q_{acc}\}$.

An MM-1QFA is defined as a 6-tuple
$A=(Q,Q_{acc},Q_{rej},|\psi_{0}\rangle,\Sigma,\{U(\sigma)\}_{\sigma\in\Sigma\cup
\{\$\}})$, where $Q,Q_{acc}\subseteq
Q,|\psi_{0}\rangle,\Sigma,\{U(\sigma)\}_{\sigma\in\Sigma\cup
\{\$\}}$ are the same as those in an MO-1QFA defined above,
$Q_{rej}\subseteq Q$ represents the set of rejecting states, and
$\$\not\in\Sigma$ is a tape symbol denoting the right end-mark.
For any input string
$x=\sigma_{1}\sigma_{2}\cdots\sigma_{m}\in\Sigma^{*}$, the
computing process is similar to that of MO-1QFAs except that after
every transition, $A$ measures its state with respect to the three
subspaces that are spanned by the three subsets $Q_{acc},
Q_{rej}$, and $Q_{non}$, respectively, where $Q_{non}=Q\setminus
(Q_{acc}\cup Q_{rej})$. In other words, the projection measurement
consists of $\{P_{a},P_{r},P_{n}\}$ where $P_{a}=\sum_{q\in
Q_{acc}}|q\rangle\langle q|$, $P_{r}=\sum_{q\in
Q_{rej}}|q\rangle\langle q|$, $P_{n}=\sum_{q\in Q\setminus
(Q_{acc}\cup Q_{rej})}|q\rangle\langle q|$. The machine stops
after the right end-mark $\$$ has been read. Of course, the
machine may also stop before reading  $\$$ if the current state of
the machine reading some $\sigma_{i}$ $(1\leq i\leq m)$ does not
contain the states of $Q_{non}$. Since the measurement is
performed after each transition with the states of $Q_{non}$ being
preserved, the accepting probability $p_{a}(x)$ and the rejecting
probability $p_{r}(x)$ are given as follows (for convenience, we
denote $\$=\sigma_{m+1}$):

\[
p_{a}(x)=\sum_{k=1}^{m+1}\|P_{a}U(\sigma_{k})\prod_{i=1}^{k-1}(P_{n}U(\sigma_{i}))|\psi_{0}\rangle\|^{2},
\]
\[
p_{r}(x)=\sum_{k=1}^{m+1}\|P_{r}U(\sigma_{k})\prod_{i=1}^{k-1}(P_{n}U(\sigma_{i}))|\psi_{0}\rangle\|^{2}.
\]

We further recall the definitions of a {\it group finite
automaton} (GFA) \cite{BP02} and a {\it one-way reversible finite
automaton} (1RFA) \cite{AF98}. A GFA is a DFA whose state
transition function, say $\delta$, satisfies that for any input
symbol $\sigma$, $\delta(\cdot,\sigma)$ is a one-to-one map on the
state set, i.e., a permutation. A 1RFA is defined as an MO-1QFA
but restricting the values of its state transition function onto
$\{0,1\}$. More specifically, a 1RFA is  a DFA whose set of
states, input alphabet, and state transition function are
$Q,\Sigma,\delta$, respectively, where $\delta$ satisfies that,
for any $q\in Q$ and any $\sigma\in \Sigma$, there is at most one
$p\in Q$ such that $\delta(p,\sigma)=q$.

Qiu \cite{Qiu07} proved that GFAs and 1RFAs are equivalent, i.e.,
any GFA can be simulated by a 1RFA and vice-versa.

\begin{Df}[{\cite{BRS07}}]
 A $k$-letter DFA $(Q,Q_{acc},q_{0},\Sigma,\gamma)$ is called a $k$-letter group finite automaton
 ($k$-letter GFA) if and only if for any string $x\in T^{k}$ the
 function $\gamma_{x}(q)=\gamma(q,x)$ is a bijection from
 $Q$ to $Q$.
\end{Df}

\begin{Rm}
When $k=1$, a $1$-letter DFA is exactly a DFA
\cite{RS97,Sal85,Yu98}, and a $1$-letter GFA is also the usual GFA
\cite{BP02}. By ${\cal L}(GFA_{k})$ and ${\cal L}(DFA_{k})$ we
denote the classes of all languages accepted by $k$-letter GFAs
and by $k$-letter DFAs, respectively. In addition, we denote
${\cal L}(GFA_{*})=\bigcup _{k=1}^{\infty}{\cal L}(GFA_{k})$ and
${\cal L}(DFA_{*})=\bigcup _{k=1}^{\infty}{\cal L}(DFA_{k})$. In
\cite{BRS07} it was shown that
\begin{equation}{\cal L}(GFA)\subset {\cal L}(GFA_{*})\subset
{\cal L}(DFA)= {\cal L}(DFA_{*}),
\end{equation}
where $\subset$ is a proper inclusion.
\end{Rm}

Now we further recall the definition of multi-letter QFAs
\cite{BRS07}.

\begin{Df} [{\cite{BRS07}}]
A $k$-letter QFA ${\cal A}$ is defined as a quintuple ${\cal
A}=(Q,Q_{acc},|\psi_{0}\rangle, \Sigma,\mu)$ where $Q$ is a set of
states, $Q_{acc}\subseteq Q$ is the set of accepting states,
$|\psi_{0}\rangle$ is the initial unit state that is a
superposition of the states in $Q$, $\Sigma$ is a finite input
alphabet, and $\mu$ is a function that assigns a unitary
transition matrix $U_{w}$ on $\mathbb{C}^{|Q|}$ for each string
$w\in (\{\Lambda\}\cup\Sigma)^{k}$, where $|Q|$ is the cardinality
of $Q$.

\end{Df}

The computation of a $k$-letter QFA ${\cal A}$ works in the same
way as the computation of an MO-1QFA, except that it applies
unitary transformations corresponding not only to the last letter
but the last $k$ letters received (like a $k$-letter DFA). When
$k=1$, it is exactly an MO-1QFA as pointed out before. According
to \cite{BRS07}, all languages accepted by $k$-letter QFAs with
bounded error are regular languages for any $k$.

Now we give the probability $P_{{\cal A}}(x)$ for $k$-letter QFA
${\cal A}=(Q,Q_{acc},|\psi_{0}\rangle, \Sigma,\mu)$ accepting any
input string $x=\sigma_{1}\sigma_{2}\cdots\sigma_{m}$. From the
definition we know that, for any $w\in
(\{\Lambda\}\cup\Sigma)^{k}$, $\mu(w)$ is a unitary matrix. In
terms of the definition of $\mu$, we can define the unitary
transition for each string
$x=\sigma_{1}\sigma_{2}\cdots\sigma_{m}\in\Sigma^{*}$. By
$\overline{\mu}$ we mean a map from $\Sigma^{*}$ to the set of all
$|Q|$-order unitary matrices. Indeed, $\overline{\mu}$ is induced
by $\mu$ in the following way. For
$x=\sigma_{1}\sigma_{2}\cdots\sigma_{m}\in\Sigma^{*}$,
\begin{equation}
\overline{\mu}(x)=\left\{\begin{array}{ll}
\mu(\Lambda^{k-1}\sigma_{1})\mu(\Lambda^{k-2}\sigma_{1}\sigma_{2})\cdots\mu(\Lambda^{k-m}x), &  {\rm if} \  m<k, \\
\mu(\Lambda^{k-1}\sigma_{1})\mu(\Lambda^{k-2}\sigma_{1}\sigma_{2})\cdots\mu(\sigma_{m-k+1}\sigma_{m-k+2}\cdots\sigma_{m}),
& {\rm if}\  m\geq k,
   \end{array}
 \right.
\end{equation}
which implies the computing process of ${\cal A}$ for input string
$x$.

As before, we identify the states in $Q$ with an orthonormal basis
of the complex Euclidean space $\mathbb{C}^{|Q|}$, and let
$P_{acc}$ denote the projector on the subspace spanned by
$Q_{acc}$. Then we define that
\begin{equation}
P_{{\cal A}}(x)=\|\langle\psi_{0}|\overline{\mu}(x)P_{acc}\|^{2}.
\end{equation}

\begin{Df} [{\cite{BRS07}}] For $k\geq 1$,
a DFA contains a {\it $C_{k}$-construction} if and only if there
are states $q_{1},q_{2},q_{3},q_{4},q_{5}$ and a string
$w=\sigma_{1}\sigma_{2}\cdots\sigma_{k}$ of length $k$ such that
$q_{2}\not= q_{5}$, and transformation function $\gamma$ satisfies
$\gamma(q_{2},\sigma_{k})=\gamma(q_{5},\sigma_{k})=q_{3}$,
$\gamma^{*}(q_{1},\sigma_{1}\cdots\sigma_{k-1})=q_{2}$ and
$\gamma^{*}(q_{4},\sigma_{1}\cdots\sigma_{k-1})=q_{5}$.

\end{Df}

In the above $C_{k}$-construction, if there exists an $m>0$ such
that $\gamma^{*}(q_{3},w^{m-1})=q_{4}$, then we call it a {\it
$D_{k}$-construction}.

\begin{Pp}  [\cite{BRS07}] If there
exists a $C_{k}$-construction in a DFA, then there also exists a
$D_{k}$-construction in this DFA.
\end{Pp}

\begin{Th}[{\cite{BRS07}}]

The following statements are equivalent:
\begin{itemize}
\item A language $L$ is in ${\cal L}(QFA_{k})$, i.e., $L$ is
accepted by a $k$-letter QFA with bounded error.

\item The minimal DFA of $L$ contains no $C_{k}$-construction.

\item $L$ is accepted by a $k$-letter GFA.
\end{itemize}

\end{Th}

From Theorem 2 we know that a language is accepted by a $k$-letter
GFA if and only if it is accepted by a $k$-letter QFA with bounded
error. For $k=1$, it was proved by Brodsky and Pippenger
\cite{BP02}.

\section*{3. Hierarchy of multi-letter QFAs and some relations}

In this section, we deal with two issues. In Subsection 3.1, we
consider the hierarchy of multi-letter QFAs and  prove that
$j$-letter QFA are strictly more powerful than $i$-letter QFAs for
$1\leq i< j$. In Subsection 3.2, we attempt to clarify the
relations between the families of languages accepted by
multi-letter QFAs and MO-1QFAs and also between those by
multi-letter QFAs and MM-QFAs.

\subsection*{3.1. Hierarchy of multi-letter QFAs}

Are $k$-letter QFAs  more powerful than $(k-1)$-letter QFAs for
$k=1,2,\ldots$? The answer is positive for $k=2$ as proved in
\cite{BRS07}. In this subsection, we demonstrate that $k$-letter
QFAs  are more powerful than $(k-1)$-letter QFAs for any $k\geq
3$.

\begin{Th}
For any $k\geq 3$, there exists a language that can be accepted by
a $k$-letter GFA but cannot be accepted by any $(k-1)$-letter GFA.

\end{Th}

\noindent\textbf{Proof.} We consider the regular language
$(a_{1}+a_{2}+\ldots+a_{k})^{*}a_{1}a_{2}\cdots a_{k-1}$ denoted
by $L_{k}$ over alphabet $\Sigma=\{a_{1},a_{2},\ldots,a_{k}\}$,
and we will prove that $L_{k}$ satisfies the theorem. First we
construct a minimal DFA for $L_{k}$ as
$A_{k}=(Q,\Sigma,q_{0},\delta,F)$ where:
\begin{itemize}
\item $Q=\{q_{0},q_{1},\ldots,q_{k-1}\}$;

\item $\Sigma=\{a_{1},a_{2},\ldots,a_{k}\}$;

\item $F=\{q_{k-1}\}$;

\item $\delta$ is defined as follows:
\begin{itemize}
\item  $\delta(q_{0},a_{1})=q_{1}$; $\delta(q_{0},a_{i})=q_{0}$
for $i=2,3,\ldots,k$;

\item $\delta(q_{1},a_{1})=q_{1}$; $\delta(q_{1},a_{2})=q_{2}$;
$\delta(q_{1},a_{i})=q_{0}$ for $i=3,4,\ldots,k$;

\item $\delta(q_{l},a_{l+1})=q_{l+1}$ and
$\delta(q_{l},a_{t})=q_{0}$ for $l=2,3,\ldots,k-1$ and
$t\in\{2,\ldots,l,l+2,l+3,\ldots,k\}$, where we denote
$q_{k}=q_{0}$.

\item $\delta(q_{i},a_{1})=q_{1}$ for $i=2,3,\ldots,k-1$.

\end{itemize}

\end{itemize}

Figure 1 depicts the DFA $A_{k}$ above described. We prove that
$A_{k}$ is a minimal DFA. It suffices to prove that, for all
states $q_{0},q_{1},\cdots,q_{k-1}$, any two different states are
distinguishable \cite{HU79}. In other words, for any $0\leq
i,j\leq k-1$ with $i\not=j$, there exists $w\in \Sigma^{*}$ such
that exactly one of $\delta^{*}(q_{i},w)$ and
$\delta^{*}(q_{j},w)$ is the accepting state $q_{k-1}$. Indeed, we
can divide it into three cases.
\begin{enumerate}

\item  $q_{i}$ and $q_{k-1}$ for $0\leq i\leq k-2$. Take
$w=\epsilon$, empty string. Then
$\delta^{*}(q_{i},\epsilon)=q_{i}$ and
$\delta^{*}(q_{k-1},\epsilon)=q_{k-1}$.

\item $q_{0}$ and $q_{l}$ for $1\leq l\leq k-2$. Take
$w=a_{l+1}a_{l+2}\cdots a_{k-1}$. Then $\delta^{*}(q_{0},w)=q_{0}$
and $\delta^{*}(q_{l},w)=q_{k-1}$.

\item $q_{i}$ and $q_{j}$ for $1\leq i<j\leq k-2$. Take
$w=a_{j+1}a_{j+2}\cdots a_{k-1}$. Then
$\delta^{*}(q_{j},w)=q_{k-1}$ and $\delta^{*}(q_{i},w)=q_{0}$.

\end{enumerate}

Therefore, we have proved that any two different states of
$q_{0},q_{1},\cdots,q_{k-1}$ are distinguishable. Consequently,
$A_{k}$ is minimal.

In fact, we can see that the number $k$ of states is minimal from
the number of equivalence classes over $\Sigma^{*}$ \cite{HU79}.
This equivalence relation $\equiv$ is defined as: for any
$w_{1},w_{2}\in\Sigma^{*}$, $w_{1}\equiv w_{2}$ iff for any
$z\in\Sigma^{*}$, either both $w_{1}z$ and $w_{2}z$ in $L_{k}$, or
neither $w_{1}z$ nor $w_{2}z$ in $L_{k}$. Then we can divide
$\Sigma^{*}$ into the following $k$ equivalence classes:
$[\epsilon], [a_{1}], [a_{1}a_{2}], \cdots, [a_{1}a_{2}\cdots
a_{k-1}]$. As a result, $k$ is the number of states of the minimal
 DFA accepting $L_{k}$.

In the state transition figure of $A_{k}$, we find  a
$C_{k-1}$-construction. In fact, set $w=a_{2}a_{3}\ldots a_{k}$.
Since $\delta(q_{0},a_{i})=q_{0}$ for $i=2,3,\ldots,k$, we get
$\delta^{*}(q_{0},a_{2}a_{3}\cdots a_{k-1})=q_{0}$ and
$\delta(q_{0},a_{k})=q_{0}$. Moreover,
$\delta(q_{i},a_{i+1})=q_{i+1}$ for $i=1,2,\ldots,k$, where we
denote $q_{k}=q_{0}$. This $C_{k-1}$-construction is better
described by Figure 2. By Theorem 2, we conclude that $A_{k}$
cannot be accepted by any $(k-1)$-letter QFA with bounded error.

\begin{center}
\setlength{\unitlength}{0.07cm}
\begin{picture}(190,76)

\put(5,40){\circle{10}\makebox(0,0){$q_{0}$}}

\put(10,40){\makebox(20,5)[c]{$a_{1}$}}

 \put(10,40){\vector(1,0){20}}

\put(5,50){\circle{11}} \put(0,50){\vector(0,-1){1}}
\put(5,58){\makebox(0,0)[c]{$a_{2},a_{3},\cdots,a_{k}$}}

\put(35,40){\circle{10}\makebox(0,0){$q_{1}$}}

\put(40,40){\makebox(20,5)[c]{$a_{2}$}}
 \put(40,40){\vector(1,0){20}}
 \put(65,40){\circle{10}\makebox(0,0){$q_{2}$}}
\put(80,40){\makebox(0,5)[c]{$a_{3}$}}\put(70,40){\vector(1,0){20}}

\put(97,37){\makebox(0,5)[c]{$\ldots$}}

\put(160,37){\makebox(0,5)[c]{$\ldots$}}

\put(35,50){\circle{11}} \put(40,50){\vector(0,-1){1}}

\put(35,58){\makebox(0,0)[c]{$a_{1}$}}

\put(105,40){\vector(1,0){20}}

 \put(130,40){\circle{10}\makebox(0,0){$q_{l}$}}

\put(135,40){\vector(1,0){20}}
\put(165,40){\makebox(20,5)[c]{$a_{k-1}$}}
 \put(165,40){\vector(1,0){20}}
 \put(193,40){\circle{12}\makebox(0,0){$q_{k-1}$}}
\put(193,40){\circle{14}}

\qbezier(9,34)(18,28)(32,34) \put(9,33){\vector(-1,1){2}}

\put(30,29){\makebox(0,0)[c]{$a_{3},a_{4},\cdots,a_{k}$}}

\qbezier(8,32)(25,15)(62,32) \put(9,31){\vector(-1,1){2}}

\put(45,20){\makebox(0,0)[c]{$a_{2},a_{4},a_{5},\cdots,a_{k}$}}

\qbezier(6,30)(50,-10)(129,34) \put(7,29){\vector(-1,1){2}}

\put(70,7){\makebox(0,0)[c]{$a_{2},\cdots,a_{l-1},a_{l},a_{l+2},\cdots,a_{k}$}}

\qbezier(5,29)(50,-35)(190,34) \put(6,28){\vector(-1,1){2}}

\put(80,-5){\makebox(0,0)[c]{$a_{2},\cdots,a_{k}$}}

\qbezier(39,44)(50,55)(62,45) \put(40,45){\vector(-1,-1){2}}
\put(52,52){\makebox(0,0)[c]{$a_{1}$}}

\qbezier(40,46)(70,80)(125,45) \put(41,47){\vector(-1,-1){2}}
\put(100,60){\makebox(0,0)[c]{$a_{1}$}}

\qbezier(41,49)(75,95)(187,45) \put(42,50){\vector(-1,-1){2}}
\put(150,65){\makebox(0,0)[c]{$a_{1}$}}

\qbezier(6,30)(50,-10)(129,34) \put(7,29){\vector(-1,1){2}}
\put(70,7){\makebox(0,0)[c]{$a_{2},\cdots,a_{l-1},a_{l},a_{l+2},\cdots,a_{k}$}}

\put(55,-20){\makebox(55,0)[c]{{\footnotesize Figure 1. A state
transition diagram of DFA $A_{k}$.}}}
\end{picture}
\end{center}

\vskip 15mm

However,  we will verify that, in the minima DFA $A_{k}$, there is
no $C_{k}$-construction. Therefore, according to Theorem 2,
$L_{k}$ can be accepted by a $k$-letter QFA with bounded error.

\vskip 25mm

\begin{center}
\setlength{\unitlength}{0.07cm}
\begin{picture}(120,50)

\put(5,40){\circle{10}\makebox(0,0){$q_{0}$}}
\put(10,40){\makebox(20,5)[c]{$a_{2}$}}
 \put(10,40){\vector(1,0){20}}
\put(35,40){\circle{10}\makebox(0,0){$q_{0}$}}
 \put(40,40){\vector(1,0){20}}

 \put(67,38){\makebox(0,5)[c]{$\ldots$}}
 \put(67,58){\makebox(0,5)[c]{$\ldots$}}

\put(75,40){\vector(1,0){20}}
 \put(101,40){\circle{12}\makebox(0,0){$q_{0}$}}
\put(75,40){\makebox(20,5)[c]{$a_{k-1}$}}

\put(5,60){\circle{10}\makebox(0,0){$q_{1}$}}
\put(10,60){\makebox(20,5)[c]{$a_{2}$}}
 \put(10,60){\vector(1,0){20}}
\put(35,60){\circle{10}\makebox(0,0){$q_{2}$}}
 \put(40,60){\vector(1,0){20}}
\put(75,60){\vector(1,0){20}}
 \put(101,60){\circle{12}\makebox(0,0){$q_{k-1}$}}
\put(75,60){\makebox(20,5)[c]{$a_{k-1}$}}

\put(107,40){\vector(3,2){10}} \put(107,60){\vector(3,-2){10}}

\put(103,37){\makebox(20,5)[c]{$a_{k}$}}
\put(103,57){\makebox(20,5)[c]{$a_{k}$}}

\put(123,50){\circle{12}\makebox(0,0){$q_{0}$}}

\put(37,60){\makebox(20,5)[c]{$a_{3}$}}
\put(37,40){\makebox(20,5)[c]{$a_{3}$}}

\put(30,25){\makebox(55,0)[c]{{\footnotesize Figure 2. A
$C_{k-1}$-construction in DFA $A_{k}$.}}}
\end{picture}
\end{center}

Now we check that there is no $C_{k}$-construction in $A_{k}$. We
prove it by contradiction. Indeed, suppose that there is a
$C_{k}$-construction depicted by Figure 3.

\vskip 20mm

\begin{center}
\setlength{\unitlength}{0.07cm}
\begin{picture}(120,50)

\put(5,40){\circle{10}\makebox(0,0){$q_{j_{1}}$}}
\put(10,40){\makebox(20,5)[c]{$\sigma_{1}$}}
 \put(10,40){\vector(1,0){20}}
\put(35,40){\circle{10}\makebox(0,0){$q_{j_{2}}$}}
 \put(40,40){\vector(1,0){20}}

 \put(67,38){\makebox(0,5)[c]{$\ldots$}}
 \put(67,58){\makebox(0,5)[c]{$\ldots$}}

\put(75,40){\vector(1,0){20}}
 \put(101,40){\circle{12}\makebox(0,0){$q_{j_{k}}$}}
\put(75,40){\makebox(20,5)[c]{$\sigma_{k-1}$}}

\put(5,60){\circle{10}\makebox(0,0){$q_{i_{1}}$}}
\put(10,60){\makebox(20,5)[c]{$\sigma_{1}$}}
 \put(10,60){\vector(1,0){20}}
\put(35,60){\circle{10}\makebox(0,0){$q_{i_{2}}$}}

\put(37,60){\makebox(20,5)[c]{$\sigma_{2}$}}
\put(37,40){\makebox(20,5)[c]{$\sigma_{2}$}}

 \put(40,60){\vector(1,0){20}}
\put(75,60){\vector(1,0){20}}
 \put(101,60){\circle{12}\makebox(0,0){$q_{i_{k}}$}}
\put(75,60){\makebox(20,5)[c]{$\sigma_{k-1}$}}

\put(107,40){\vector(3,2){10}} \put(107,60){\vector(3,-2){10}}

\put(103,37){\makebox(20,5)[c]{$\sigma_{k}$}}
\put(103,57){\makebox(20,5)[c]{$\sigma_{k}$}}

\put(123,50){\circle{12}\makebox(0,0){$q$}}

\put(25,25){\makebox(55,0)[c]{{\footnotesize Figure 3. A supposed
$C_{k}$-construction. }}}
\end{picture}
\end{center}

We divide the proof into the following three cases.
\begin{enumerate}

\item $q=q_{0}$.

\begin{itemize}

\item $\sigma_{k}=a_{1}$: It is impossible since $q_{0}$ cannot be
accessed by inputting $a_{1}$.

\item $\sigma_{k}=a_{2}$: In this case, $q_{i_{k}},q_{j_{k}}\in
\{q_{0},q_{2},q_{3},\ldots,q_{k-1}\}$.

If one of $q_{i_{k}},q_{j_{k}}$, say $q_{i_{k}}$ is $q_{0}$, and
the other one $q_{j_{k}}$ belongs to
$\{q_{2},q_{3},\ldots,q_{k-1}\}$, then $\sigma_{k-1}=a_{j_{k}}$
where $j_{k}\geq 2$. Thus, $q_{j_{k-1}}=q_{j_{k}-1}$ and
$q_{i_{k-1}}=q_{0}$. In succession, we find that $q_{i_{t}}=q_{0},
q_{j_{t}}=q_{1}$ for some $2\leq t\leq k$. However, there is no
$\sigma\in\Sigma$ leading to $q_{0}$ and $q_{1}$ simultaneously.
Therefore, it is impossible.

If $q_{i_{k}},q_{j_{k}}\in \{q_{2},q_{3},\ldots,q_{k-1}\}$, then
the above case shows that this is impossible either.

Consequently, this case does not exist.

\item  $\sigma_{k}=a_{s}$ for $3\leq s\leq k$:

These cases can be similarly verified as above, and we leave the
details out here.

\end{itemize}

\item $q=q_{1}$

\begin{itemize}

\item $\sigma_{k}=a_{1}$: In this case, $q_{i_{k}},q_{j_{k}}\in
\{q_{0},q_{1},q_{2},q_{3},\ldots,q_{k-1}\}$. Similar to the above
proof.

\item $\sigma_{k}=a_{s}$ for $2\leq s\leq k$: It is clearly
impossible.

\end{itemize}

\item $q=q_{s}$ for $2\leq s\leq k$:

For any  $2\leq s\leq k$, there is no $\sigma\in\Sigma$ and two
different states $p_{1}\not=p_{2}$ such that
$\delta(p_{1},\sigma)=\delta(p_{2},\sigma)=q_{s}$. Consequently,
there does not exist such a $C_{k}$-construction.

\end{enumerate}

Hence, there does not exist a $C_{k}$-construction in $A_{k}$, and
therefore, by Theorem 2, $L_{k}$ can be accepted by a $k$-letter
GFA.

\qed\\

From Theorem 2 and Theorem 3 we have the following corollary.

\begin{Co}

For $k\geq 2$, ${\cal L}(QFA_{k-1})\subset {\cal L}(QFA_{k})$,
where the inclusion is proper.

\end{Co}

\subsection*{3.2. Comparison of multi-letter QFAs with others}

In this subsection, we try to compare the relations between the
families of languages accepted by multi-letter QFAs and MO-1QFAs
and also between those by multi-letter QFAs and MM-QFAs. First we
recall the definition of {\it forbidden construction} in a DFA
\cite{AF98}.

In a DFA, a forbidden construction means that there exist string
$x$ and states $p_{1}$ and $p_{2}$, $p_{1}\not=p_{2}$, such that
$\delta^{*}(p_{1},x)=p_{2}$ and $\delta^{*}(p_{2},x)=p_{2}$, where
$p_{2}$ is neither ``all-accepting" state, nor ``all-rejecting". A
state $p$ is neither ``all-accepting" state, nor ``all-rejecting"
whenever there exist $w_{1},w_{2}\in\Sigma^{*}$ such that exactly
one of $\delta^{*}(p,w_{1})$ and $\delta^{*}(p,w_{2})$ is an
accepting state.

\begin{Rm}
Ambainis and Freivalds \cite{AF98}  presented a forbidden
construction and showed that, if the minimal DFA for accepting a
regular language does not contain a forbidden construction, then
this language can be accepted by a one-way reversible finite
automaton. In \cite{Qiu07}, Qiu proved that one-way reversible
finite automata are also GFAs and vice versa. Also, Ambainis and
Freivalds \cite{AF98} proved that a regular language is accepted
by an MM-1QFA with bounded error and with probability over
$\frac{7}{9}$ if and only if this language is accepted by a 1RFA
and thus by a GFA as well.
\end{Rm}

Next we verify that a forbidden construction implies a
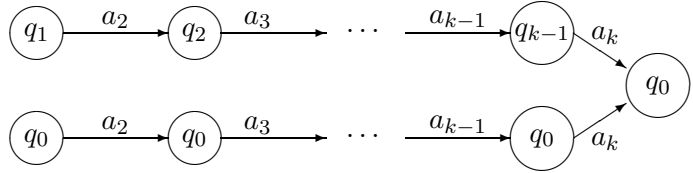
$C_{1}$-construction.

\begin{Pp}

In a DFA, if there exists a forbidden construction, then there
also exists a $C_{1}$-construction.

\end{Pp}

\noindent\textbf{Proof.} Let $A=(Q,Q_{acc},q_{0},\Sigma,\delta)$
be a DFA. Suppose that  there is a forbidden construction, that
is, there are states $p_{1},p_{2}$ and $x\in\Sigma^{*}$ satisfying
$\delta(p_{1},x)=p_{2}$ and $\delta(p_{2},x)=p_{2}$. Suppose that
$x=\sigma_{1}\sigma_{2}\ldots\sigma_{k}$. Then there are states
$q_{1},q_{2},\ldots,q_{k}$ and $r_{1},r_{2},\ldots,r_{k}$ with
$q_{k}=p_{2}=r_{k}$ such that $\delta(p_{1},\sigma_{1})=q_{1}$,
$\delta(q_{k},\sigma_{1})=r_{1}$,
$\delta(q_{i},\sigma_{i+1})=q_{i+1}$, where $i=1,2,\ldots,k-1$.
This relation can be described by Figure 4.

\begin{center}
\setlength{\unitlength}{0.07cm}
\begin{picture}(180,50)

\put(5,40){\circle{10}\makebox(0,0){$p_{1}$}}
\put(8,40){\makebox(20,5)[c]{$\sigma_{1}$}}
 \put(10,40){\vector(1,0){15}}
\put(30,40){\circle{10}\makebox(0,0){$q_{1}$}}
\put(32,40){\makebox(20,5)[c]{$\sigma_{2}$}}
 \put(35,40){\vector(1,0){15}}
\put(55,40){\circle{10}\makebox(0,0){$q_{2}$}}
\put(60,40){\vector(1,0){15}}

 \put(80,38){\makebox(0,5)[c]{$\ldots$}}
\put(85,40){\makebox(20,5)[c]{$\sigma_{k}$}}
\put(85,40){\vector(1,0){15}}

 \put(105,40){\circle{10}\makebox(0,0){$q_{k}$}}

\put(110,40){\makebox(20,5)[c]{$\sigma_{1}$}}
 \put(110,40){\vector(1,0){15}}
\put(130,40){\circle{10}\makebox(0,0){$r_{1}$}}
\put(135,40){\vector(1,0){15}}
\put(155,38){\makebox(0,5)[c]{$\ldots$}}

\put(160,40){\makebox(20,5)[c]{$\sigma_{k}$}}
\put(160,40){\vector(1,0){15}}
 \put(180,40){\circle{10}\makebox(0,0){$r_{k}$}}
\put(50,20){\makebox(55,0)[c]{{\footnotesize Figure 4. A relation
diagram where $q_{k}=p_{2}=r_{k}$. }}}
\end{picture}
\end{center}

Since $p_{1}\not=p_{2}=q_{k}$ but $q_{k}=r_{k}$, there exists
$q_{i}=r_{i}$ but $q_{i-1}\not= r_{i-1}$. Therefore, we have
$\delta(q_{i-1},\sigma_{i})=q_{i}=r_{i}$ and
$\delta(r_{i-1},\sigma_{i})=r_{i}=r_{i}$, which is a
$C_{1}$-construction.

\qed\\

By Remark 2 and Theorem 2 we obtain the following corollary.

\begin{Pp}

The minimal DFA accepting a regular language $L$ does not contain
$C_{1}$-construction if and only if $L$ can be accepted by  an
MM-1QFA with bounded error and with probability over
$\frac{7}{9}$.

\end{Pp}

\begin{proof}

If the minimal DFA accepting a regular language $L$ does not
contain $C_{1}$-construction, then, by Theorem 2 we obtain that
$L$ can be accepted by a GFA. Therefore, by Remark 2, $L$ can be
accepted by an MM-1QFA with bounded error and with probability
over $\frac{7}{9}$.

On the other hand, if $L$ is accepted by an MM-1QFA with bounded
error and with probability over $\frac{7}{9}$, then, with Remark 2
we know that $L$ can be accepted by a GFA. By Theorem 2, the
minimal DFA accepting $L$ does not contain $C_{1}$-construction.

\end{proof}

Next, we present  a few examples to show that ${\cal L}(QFA_{*})$
is still a proper subset of all regular languages. Let us first
show an example of regular language that can be accepted by an
MM-1QFA but not by any multi-letter QFA.

\begin{Emp} The language $a^{*}b^{*}$ can be accepted by an
MM-1QFA \cite{AF98} but it cannot be accepted by any $k$-letter
QFA. Indeed, we can describe the minimal DFA $M$ for accepting
$a^{*}b^{*}$ by Figure 5. In addition, from this figure we can
find that there exists a $C_{k}$-construction for any $k\geq 2$,
which is visualized by Figure 6.

\end{Emp}

\vskip 20mm

\begin{center}
\setlength{\unitlength}{0.07cm}
\begin{picture}(80,50)

\put(10,60){\circle{12}\makebox(0,0){$q_{0}$}}
\put(10,60){\circle{10}\makebox(0,0){$q_{0}$}}
\put(10,40){\circle{10}\makebox(0,0){$q_{2}$}}

 \put(15,60){\vector(3,-2){10}}

\put(10,57){\makebox(20,5)[c]{$b$}}

\put(25,45){\vector(-3,-2){10}}

\put(10,43){\makebox(20,5)[c]{$a$}}

\put(30,50){\circle{12}\makebox(0,0){$q_{1}$}}
\put(30,50){\circle{10}\makebox(0,0){$q_{1}$}}

\put(-1,60){\circle{11}} \put(3,57){\vector(1,1){1}}

\put(-18,58){\makebox(20,5)[c]{$a$}}

\put(40,55){\circle{11}} \put(37,51){\vector(-1,1){1}}
\put(37,55){\makebox(20,5)[c]{$b$}}

\put(1,35){\circle{11}} \put(6,35){\vector(0,1){1}}
\put(-16,33){\makebox(20,5)[c]{$a$}}

\put(16,32){\circle{11}} \put(11,33){\vector(0,1){1}}
\put(11,25){\makebox(20,5)[c]{$b$}}

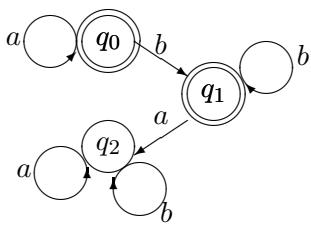
\put(5,15){\makebox(55,0)[c]{{\footnotesize Figure 5. A state
transition diagram of DFA $M$ accepting $a^{*}b^{*}$. }}}
\end{picture}
\end{center}

\vskip 20mm

\begin{center}
\setlength{\unitlength}{0.07cm}
\begin{picture}(160,50)

 \put(40,40){\makebox(20,5)[c]{$a$}}
\put(35,40){\circle{10}\makebox(0,0){$q_{1}$}}
 \put(40,40){\vector(1,0){20}}

 \put(67,40){\circle{10}\makebox(0,0){$q_{2}$}}
 \put(67,60){\circle{10}\makebox(0,0){$q_{0}$}}

\put(75,40){\vector(1,0){20}}
 \put(101,40){\circle{10}\makebox(0,0){$q_{2}$}}
\put(75,40){\makebox(20,5)[c]{$b^{k-1}$}}

 \put(40,60){\makebox(20,5)[c]{$a^{k-1}$}}

\put(35,60){\circle{10}\makebox(0,0){$q_{0}$}}
 \put(40,60){\vector(1,0){20}}
\put(75,60){\vector(1,0){20}}
 \put(101,60){\circle{10}\makebox(0,0){$q_{1}$}}
\put(75,60){\makebox(20,5)[c]{$b$}}

\put(107,40){\vector(3,2){10}} \put(107,60){\vector(3,-2){10}}

\put(103,37){\makebox(20,5)[c]{$a$}}
\put(103,57){\makebox(20,5)[c]{$a$}}

\put(123,50){\circle{10}\makebox(0,0){$q_{2}$}}

\put(45,20){\makebox(55,0)[c]{{\footnotesize Figure 6.  A
$C_{k}$-construction in Figure 6.}}}
\end{picture}
\end{center}

Next we provide another example which demonstrates that there
exist regular languages  acceptable  neither by MM-1QFAs nor by
multi-letter QFAs with bounded error. However, we need a result
from \cite{BRS07}.

\begin{Df} [\cite{BRS07}]
A DFA  with state transition function $\delta$ is said to contain
an F-construction if and only if there are non-empty words $t,z\in
\Sigma^{+}$ and two distinct states $q_{1},q_{2}\in Q$ such that
$\delta^{*}(q_{1},z)=\delta^{*}(q_{2},z)=q_{2}$,
$\delta^{*}(q_{1},t)=q_{1}$, $\delta^{*}(q_{2},t)=q_{2}$.

\end{Df}

\begin{Pp} [\cite{BRS07}]
A language $L$ can be accepted by a multi-letter QFA with bounded
error if and only if the minimal DFA of $L$ does not contain any
F-construction.

\end{Pp}

\begin{Emp} We use an example from \cite{AKV01}. Let $L$ be the language consisting of all words that
start with any number of letters $a$ and after first letter $b$
(if there is one) there is an odd number of letters $a$. The
minimal DFA $G$ accepting $L$ is depicted by Figure 7. As proved
by Ambainis {\it et al} \cite{AKV01}, $L$ cannot be accepted by
MM-1QFAs with bounded error. Indeed, $L$  cannot be accepted by
any multi-letter QFA, either. Because there exists an
F-construction (Figure 8) in the minimal DFA $G$ (Figure 7), we
get the result.

\end{Emp}

\vskip 20mm

  \begin{center}
  \begin{Large}
\special{em:linewidth 0.4pt} \unitlength 1mm \linethickness{0.4pt}
\begin{picture}(70.00,47.00)
\emline{50.00}{47.00}{1}{51.59}{46.82}{2}
\emline{51.59}{46.82}{3}{53.09}{46.28}{4}
\emline{53.09}{46.28}{5}{54.43}{45.42}{6}
\emline{54.43}{45.42}{7}{55.54}{44.27}{8}
\emline{55.54}{44.27}{9}{56.37}{42.91}{10}
\emline{56.37}{42.91}{11}{56.86}{41.39}{12}
\emline{56.86}{41.39}{13}{57.00}{39.80}{14}
\emline{57.00}{39.80}{15}{56.77}{38.22}{16}
\emline{56.77}{38.22}{17}{56.19}{36.73}{18}
\emline{56.19}{36.73}{19}{55.29}{35.42}{20}
\emline{55.29}{35.42}{21}{54.11}{34.34}{22}
\emline{54.11}{34.34}{23}{52.72}{33.55}{24}
\emline{52.72}{33.55}{25}{51.19}{33.10}{26}
\emline{51.19}{33.10}{27}{49.60}{33.01}{28}
\emline{49.60}{33.01}{29}{48.03}{33.28}{30}
\emline{48.03}{33.28}{31}{46.56}{33.90}{32}
\emline{46.56}{33.90}{33}{45.27}{34.84}{34}
\emline{45.27}{34.84}{35}{44.22}{36.05}{36}
\emline{44.22}{36.05}{37}{43.48}{37.46}{38}
\emline{43.48}{37.46}{39}{43.07}{39.00}{40}
\emline{43.07}{39.00}{41}{43.03}{40.60}{42}
\emline{43.03}{40.60}{43}{43.34}{42.16}{44}
\emline{43.34}{42.16}{45}{44.01}{43.61}{46}
\emline{44.01}{43.61}{47}{44.98}{44.88}{48}
\emline{44.98}{44.88}{49}{46.22}{45.89}{50}
\emline{46.22}{45.89}{51}{47.65}{46.59}{52}
\emline{47.65}{46.59}{53}{50.00}{47.00}{54}
\put(50.00,40.00){\circle{10.00}}
\put(50.00,40.00){\makebox(0,0)[cc]{$q_1$}}
\emline{20.00}{17.00}{55}{21.59}{16.82}{56}
\emline{21.59}{16.82}{57}{23.09}{16.28}{58}
\emline{23.09}{16.28}{59}{24.43}{15.42}{60}
\emline{24.43}{15.42}{61}{25.54}{14.27}{62}
\emline{25.54}{14.27}{63}{26.37}{12.91}{64}
\emline{26.37}{12.91}{65}{26.86}{11.39}{66}
\emline{26.86}{11.39}{67}{27.00}{9.80}{68}
\emline{27.00}{9.80}{69}{26.77}{8.22}{70}
\emline{26.77}{8.22}{71}{26.19}{6.73}{72}
\emline{26.19}{6.73}{73}{25.29}{5.42}{74}
\emline{25.29}{5.42}{75}{24.11}{4.34}{76}
\emline{24.11}{4.34}{77}{22.72}{3.55}{78}
\emline{22.72}{3.55}{79}{21.19}{3.10}{80}
\emline{21.19}{3.10}{81}{19.60}{3.01}{82}
\emline{19.60}{3.01}{83}{18.03}{3.28}{84}
\emline{18.03}{3.28}{85}{16.56}{3.90}{86}
\emline{16.56}{3.90}{87}{15.27}{4.84}{88}
\emline{15.27}{4.84}{89}{14.22}{6.05}{90}
\emline{14.22}{6.05}{91}{13.48}{7.46}{92}
\emline{13.48}{7.46}{93}{13.07}{9.00}{94}
\emline{13.07}{9.00}{95}{13.03}{10.60}{96}
\emline{13.03}{10.60}{97}{13.34}{12.16}{98}
\emline{13.34}{12.16}{99}{14.01}{13.61}{100}
\emline{14.01}{13.61}{101}{14.98}{14.88}{102}
\emline{14.98}{14.88}{103}{16.22}{15.89}{104}
\emline{16.22}{15.89}{105}{17.65}{16.59}{106}
\emline{17.65}{16.59}{107}{20.00}{17.00}{108}
\put(20.00,10.00){\circle{10.00}}
\put(50.00,10.00){\circle{10.00}}
\put(26.33,7.00){\vector(-4,1){0.2}}
\emline{46.00}{7.00}{109}{43.58}{6.42}{110}
\emline{43.58}{6.42}{111}{41.15}{6.00}{112}
\emline{41.15}{6.00}{113}{38.70}{5.75}{114}
\emline{38.70}{5.75}{115}{33.79}{5.75}{116}
\emline{33.79}{5.75}{117}{31.31}{6.00}{118}
\emline{31.31}{6.00}{119}{28.83}{6.42}{120}
\emline{28.83}{6.42}{121}{26.33}{7.00}{122}
\put(46.00,13.00){\vector(4,-1){0.2}}
\emline{26.33}{13.00}{123}{28.97}{13.58}{124}
\emline{28.97}{13.58}{125}{31.56}{14.00}{126}
\emline{31.56}{14.00}{127}{34.10}{14.25}{128}
\emline{34.10}{14.25}{129}{39.02}{14.25}{130}
\emline{39.02}{14.25}{131}{41.40}{14.00}{132}
\emline{41.40}{14.00}{133}{43.72}{13.58}{134}
\emline{43.72}{13.58}{135}{46.00}{13.00}{136}
\put(20.00,10.00){\makebox(0,0)[cc]{$q_3$}}
\put(50.00,10.00){\makebox(0,0)[cc]{$q_2$}}
\put(54.00,7.00){\vector(-4,-1){0.2}}
\emline{54.00}{13.00}{137}{56.24}{12.55}{138}
\emline{56.24}{12.55}{139}{58.11}{12.09}{140}
\emline{58.11}{12.09}{141}{59.62}{11.64}{142}
\emline{59.62}{11.64}{143}{60.76}{11.18}{144}
\emline{60.76}{11.18}{145}{61.53}{10.73}{146}
\emline{61.53}{10.73}{147}{61.93}{10.27}{148}
\emline{61.93}{10.27}{149}{61.97}{9.82}{150}
\emline{61.97}{9.82}{151}{61.64}{9.36}{152}
\emline{61.64}{9.36}{153}{60.94}{8.91}{154}
\emline{60.94}{8.91}{155}{59.88}{8.45}{156}
\emline{59.88}{8.45}{157}{58.44}{8.00}{158}
\emline{58.44}{8.00}{159}{56.64}{7.55}{160}
\emline{56.64}{7.55}{161}{54.00}{7.00}{162}
\put(13.67,7.00){\vector(4,-1){0.2}}
\emline{13.67}{13.00}{163}{11.46}{12.48}{164}
\emline{11.46}{12.48}{165}{9.67}{11.97}{166}
\emline{9.67}{11.97}{167}{8.30}{11.45}{168}
\emline{8.30}{11.45}{169}{7.34}{10.93}{170}
\emline{7.34}{10.93}{171}{6.80}{10.41}{172}
\emline{6.80}{10.41}{173}{6.67}{9.90}{174}
\emline{6.67}{9.90}{175}{6.97}{9.38}{176}
\emline{6.97}{9.38}{177}{7.67}{8.86}{178}
\emline{7.67}{8.86}{179}{8.80}{8.34}{180}
\emline{8.80}{8.34}{181}{10.34}{7.83}{182}
\emline{10.34}{7.83}{183}{13.67}{7.00}{184}
\put(53.00,14.00){\vector(-1,-3){0.2}}
\emline{53.00}{33.67}{185}{53.84}{31.36}{186}
\emline{53.84}{31.36}{187}{54.45}{29.04}{188}
\emline{54.45}{29.04}{189}{54.84}{26.72}{190}
\emline{54.84}{26.72}{191}{55.00}{24.38}{192}
\emline{55.00}{24.38}{193}{54.93}{22.04}{194}
\emline{54.93}{22.04}{195}{54.63}{19.69}{196}
\emline{54.63}{19.69}{197}{54.11}{17.32}{198}
\emline{54.11}{17.32}{199}{53.00}{14.00}{200}
\put(57.00,25.00){\makebox(0,0)[cc]{$b$}}
\put(58.00,15.00){\makebox(0,0)[cc]{$b$}}
\put(35.00,17.00){\makebox(0,0)[cc]{$a$}}
\put(35.00,8.00){\makebox(0,0)[cc]{$a$}}
\put(10.00,14.67){\makebox(0,0)[cc]{$b$}}
\put(56.33,43.00){\vector(-1,0){0.2}}
\emline{66.33}{43.00}{201}{56.33}{43.00}{202}
\put(43.67,37.00){\vector(4,-1){0.2}}
\emline{43.67}{43.00}{203}{41.46}{42.48}{204}
\emline{41.46}{42.48}{205}{39.67}{41.97}{206}
\emline{39.67}{41.97}{207}{38.30}{41.45}{208}
\emline{38.30}{41.45}{209}{37.34}{40.93}{210}
\emline{37.34}{40.93}{211}{36.80}{40.41}{212}
\emline{36.80}{40.41}{213}{36.67}{39.90}{214}
\emline{36.67}{39.90}{215}{36.97}{39.38}{216}
\emline{36.97}{39.38}{217}{37.67}{38.86}{218}
\emline{37.67}{38.86}{219}{38.80}{38.34}{220}
\emline{38.80}{38.34}{221}{40.34}{37.83}{222}
\emline{40.34}{37.83}{223}{43.67}{37.00}{224}
\put(40.00,44.33){\makebox(0,0)[cc]{$a$}}

\put(10,-5){\makebox(55,0)[c]{{\footnotesize Figure 7. Automaton
$G$.  }}}

\end{picture}
\end{Large}
\end{center}  \label{Bilde2a}

\vskip 50mm

\begin{center}
\setlength{\unitlength}{0.07cm}
\begin{picture}(60,50)

\put(-13,50){\circle{12}\makebox(0,0){$q_{1}$}}
\put(-20,59){\circle{11}} \put(-18,54.5){\vector(3,2){1}}
\put(-23,53){\makebox(0,0){$aa$}} \put(-7,50){\vector(1,0){30}}
\put(10,53){\makebox(0,0){$b$}}
\put(30,49.5){\circle{12}\makebox(0,0){$q_{2}$}}

\put(40,55){\circle{11}} \put(37,51){\vector(-1,1){1}}
\put(37,41){\circle{11}} \put(36,46){\vector(-1,0){1}}
\put(43,36){\makebox(0,0){$b$}}\put(42,62){\makebox(0,0){$aa$}}

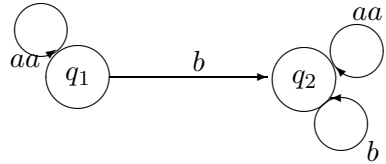
\put(0,28){\makebox(55,0)[c]{{\footnotesize Figure 8. An
F-construction in the minimal DFA  $G$.  }}}
\end{picture}
\end{center}

In conclusion, we can describe the relations between the families
of languages accepted by MO-1QFAs, MM-1QFAs, and multi-letter
QFAs, denoted by ${\cal L}(MO)$, ${\cal L}(MM)$, and ${\cal
L}(QFA_{*})$, respectively. We recall that the language
$(a+b)^{*}b$ is accepted with no error by a 2-letter QFA but
cannot be accepted by any MM-1QFA with bounded error, while
$a^{*}b^{*}$ is accepted by an MM-1QFA but cannot be accepted by
any multi-letter QFA. Therefore, both ${\cal L}(MM)\backslash
{\cal L}(QFA_{*})\not=\emptyset$  and  $ {\cal
L}(QFA_{*})\backslash {\cal L}(MM)\not=\emptyset$ hold.
Furthermore, we have that ${\cal L}(MO)\subseteq {\cal L}(MM)\cap
{\cal L}(QFA_{*})$, where $\subseteq$ may be proper. However, by
Example 2, we have known that ${\cal L}(MM)\cup {\cal L}(QFA_{*})$
still is a proper subset of all regular languages.

\section*{4. Determining the equivalence between multi-letter
quantum finite automata}

Determining whether or not two one-way (probabilistic, quantum)
finite automata and sequential machines are equivalent is of
importance and has been well studied
\cite{Paz71,Tze92,MC00,LQ06,LQ08}. Concerning multi-letter QFAs,
this issue is much more complicated and a new technique is needed.
Here, we consider only the case of unary languages, i.e., the
input alphabet having one element.

Our goal is to deal with the decidability of equivalence of unary
multi-letter QFAs. More specifically, for any given $k_{1}$-letter
QFA ${\cal A}_{1}$ and $k_{2}$-letter QFA ${\cal A}_{2}$ over the
same input alphabet $\Sigma=\{\sigma\}$, our purpose is to
determine whether or not they are equivalent.

For a $k$-letter QFA ${\cal A}=(Q,Q_{acc},|\psi_{0}\rangle,
\Sigma,\mu)$, we recall the probability $P_{{\cal A}}(x)$ for
${\cal A}$ accepting  input string
$x=\sigma_{1}\sigma_{2}\cdots\sigma_{m}$ and the definition of
$\overline{\mu}(x)$ as follows:
\begin{equation}
\overline{\mu}(x)=\left\{\begin{array}{ll}
\mu(\Lambda^{k-1}\sigma_{1})\mu(\Lambda^{k-2}\sigma_{1}\sigma_{2})\cdots\mu(\Lambda^{k-m}x), &  {\rm if} \  m<k, \\
\mu(\Lambda^{k-1}\sigma_{1})\mu(\Lambda^{k-2}\sigma_{1}\sigma_{2})\cdots\mu(\sigma_{m-k+1}\sigma_{m-k+2}\cdots\sigma_{m}),
& {\rm if}\  m\geq k,
   \end{array}
 \right.
\end{equation}
and  then
\begin{equation}
P_{{\cal A}}(x)=\|\langle\psi_{0}|\overline{\mu}(x)P_{acc}\|^{2}.
\end{equation}

We give the definition of equivalence between two multi-letter
QFAs.

\begin{Df}
A $k_{1}$-letter QFA ${\cal A}_1$ and another $k_{2}$-letter QFA
${\cal A}_2$  over the same input alphabet $\Sigma$ are said to be
equivalent (resp. $t$-equivalent) if $P_{{\cal A}_1}(w)=P_{{\cal
A}_2}(w)$ for any $w\in \Sigma^{*} $ (resp. for any input string
$w$ with $ |w|\leq t$).
\end{Df}

Before we present a method for determining the equivalence between
multi-letter QFAs over the same unary alphabet, we prove a useful
lemma that is helpful to the main result. We recall the definition
of tensor product of matrices \cite{Gru99}. For $m\times n$ matrix
$A=\left[\begin{array}{cccc} a_{11}&a_{12}&\cdots&a_{1n}\\
a_{21}&a_{22}&\cdots&a_{2n}\\
\vdots&\vdots&\cdots&\vdots\\
a_{m1}&a_{m2}&\cdots&a_{mn} \end{array} \right]$ and $p\times q$
matrix $B$, their tensor product $A\otimes B$ is an $mp \times nq$
matrix defined as
\[A\otimes B=\left[\begin{array}{cccc} a_{11}B&a_{12}B&\cdots&a_{1n}B\\
a_{21}B&a_{22}B&\cdots&a_{2n}B\\
\vdots&\vdots&\cdots&\vdots\\
a_{m1}B&a_{m2}B&\cdots&a_{mn}B \end{array} \right].\]

A basic property of tensor product is that, for any $m\times n$
matrix $A$, $p\times q$ matrix $B$, $n\times o$ matrix $C$, and
$q\times r$ matrix $D$,
\[
(A\otimes B)(C\otimes D)=(AC)\otimes(BD).
\]

Now we present the crucial lemma.

\begin{Lm} \label{Lm}
Let $\{U_{1},U_{2},\ldots,U_{k}\}$ be a finite set of $n\times n$
unitary matrices, and let $\mathbb{M}_{n^{2}}$ denote the linear
space consisting of all $n^{2}\times n^{2}$ complex square
matrices. Denote
\[
H^{(i)}=span\{(U_{1}U_{2}\cdots U_{k})\otimes
(U_{1}^{*}U_{2}^{*}\cdots U_{k}^{*}),\ldots,(U_{1}U_{2}\cdots
U_{k}^{i})\otimes (U_{1}^{*}U_{2}^{*}\cdots (U_{k}^{i})^{*})\}
\]
for $i=1,2,\cdots$, where, for any subset $A$ of
$\mathbb{M}_{n^{2}}$, $span A$ denotes the minimal subspace
spanned by $A$, and $*$ denotes the conjugate operation. Then,
there exists an $i_{0}\leq n^{4}$ such that
\begin{eqnarray}
H^{(i_{0})}&=&H^{(i_{0}+t)}\label{Lm2}
\end{eqnarray}
for any $t\geq 0$.

\end{Lm}

\begin{proof}
Let $\dim (S)$ denote the dimension of subspace $S$. Due to
\[
H^{(i)}\subseteq H^{(i+1)}\subseteq \mathbb{M}_{n^{2}},
\]
for any $i\geq 1$, we have
$1\leq\dim(H^{(1)})\leq\dim(H^{(2)})\leq\cdots\leq
\dim(H^{(n^{4}+1)})\leq n^{4}$. Therefore, we obtain that there
exists an $i_{0}\leq n^{4}$ such that $
H^{(i_{0})}=H^{(i_{0}+1)}$. Next we prove by induction that Eq.
(\ref{Lm2}) holds for $t\geq 0$. First, we have known that it
holds for $t=0,1$. Suppose that it holds for $t=i\geq 1$, i.e., $
H^{(i_{0})}=H^{(i_{0}+i)}$. This implies that $
H^{(i_{0})}=H^{(i_{0}+1)}=\cdots=H^{(i_{0}+i)}$. Our purpose is to
show that it holds for $t=i+1$, i.e.,
$H^{(i_{0})}=H^{(i_{0}+i+1)}$. Indeed, we have
\begin{eqnarray}
&&\nonumber (U_{1}U_{2}\cdots U_{k}^{i_{0}+i+1})\otimes
(U_{1}^{*}U_{2}^{*}\cdots
(U_{k}^{i_{0}+i+1})^{*})\\
&=&\nonumber \left[(U_{1}U_{2}\cdots U_{k}^{i_{0}+i})\otimes
(U_{1}^{*}U_{2}^{*}\cdots
(U_{k}^{i_{0}+i})^{*})\right](U_{k}\otimes
U_{k}^{*})\\
&=&\sum_{j=1}^{i_{0}}c_{j}[(U_{1}U_{2}\cdots U_{k}^{j})\otimes
(U_{1}^{*}U_{2}^{*}\cdots (U_{k}^{j})^{*})] (U_{k}\otimes
U_{k}^{*}) \label{Lm1}\\
&=&\sum_{j=1}^{i_{0}}c_{j} (U_{1}U_{2}\cdots U_{k}^{j+1})\otimes
(U_{1}^{*}U_{2}^{*}\cdots (U_{k}^{j+1})^{*})
\end{eqnarray}
where (\ref{Lm1}) is due to the assumption
$H^{(i_{0})}=H^{(i_{0}+i)}$. Therefore, $$(U_{1}U_{2}\cdots
U_{k}^{i_{0}+i+1})\otimes (U_{1}^{*}U_{2}^{*}\cdots
(U_{k}^{i_{0}+i+1})^{*})\in H^{(i_{0}+1)}=H^{(i_{0}+i)}.$$
Consequently, $H^{(i_{0}+i+1)}=H^{(i_{0}+i)}$. Again, by the
assumption of induction $H^{(i_{0})}=H^{(i_{0}+i)}$, we obtain
that $H^{(i_{0}+i+1)}=H^{(i_{0})}$. Therefore, (\ref{Lm2}) holds
for any $t\geq 0$.

\end{proof}

Now we are ready to present the main theorem regarding the
equivalence of multi-letter QFAs.

\begin{Th} \label{MainTH}

For $\Sigma=\{\sigma\}$, a $k_{1}$-letter QFA ${\cal
A}_1=(Q_{1},Q_{acc,1},|\psi_{0}^{(1)}\rangle, \Sigma,\mu_{1})$ and
another $k_{2}$-letter QFA ${\cal
A}_2=(Q_{2},Q_{acc,2},|\psi_{0}^{(2)}\rangle, \Sigma,\mu_{2})$ are
equivalent if and only if they are
$(n_{1}+n_{2})^{4}+k-1$-equivalent, where $n_{i}$ is the number of
states of $Q_{i}$, $i=1,2$, $k=\max(k_{1},k_{2})$, with
$k_{1},k_{2}\geq 1$.

\end{Th}

\noindent\textbf{Proof.} Let $P_{acc,1}$ and $P_{acc,2}$ denote
the projections on the subspaces spanned by $Q_{acc,1}$ and
$Q_{acc,2}$, respectively. For any string $x\in\Sigma^{*}$, we set
$\overline{\mu}(x)=\overline{\mu}_{1}(x)\oplus\overline{\mu}_{2}(x)$
and $P_{acc}=P_{acc,1}\oplus P_{acc,2}$, $Q_{acc}=Q_{acc,1}\oplus
Q_{acc,2}$, where $\oplus$ denotes the direct sum operation of any
two matrices. More precisely, for any $m_{1}\times n_{1}$ matrix
$A$ and $m_{2}\times n_{2}$ matrix $B$, $A\oplus B$ is defined as
$A\oplus
B=\left[\begin{array}{ll} A&0\\
0&B \end{array}\right]$, an $(m_{1}+m_{2})\times (n_{1}+n_{2})$
matrix.

In addition, we denote
$|\eta_{1}\rangle=|\psi_{0}^{(1)}\rangle\oplus {\bf 0_{2}}$ and
$|\eta_{2}\rangle={\bf 0_{2}}\oplus |\psi_{0}^{(2)}\rangle$, where
${\bf 0_{1}}$ and ${\bf 0_{2}}$ represent column zero vectors of
$n_{1}$ and $n_{2}$ dimensions, respectively. Then, for any string
$x\in \Sigma^{*}$,
\begin{equation}
P_{\eta_{1}}(x)=\|\langle\eta_{1}|\overline{\mu}(x)P_{acc}\|^{2}
\end{equation}
and
\begin{equation}
P_{\eta_{2}}(x)=\|\langle\eta_{2}| \overline{\mu}
(x)P_{acc}\|^{2}.
\end{equation}

Indeed, we further have that
\begin{eqnarray}
P_{\eta_{1}}(x)&=& \nonumber \|\langle\eta_{1}|\overline{\mu}(x)P_{acc}\|^{2}\\
&=&\nonumber\langle\eta_{1}|\overline{\mu}(x)P_{acc}P_{acc}^{\dagger}\overline{\mu}(x)^{\dagger}|\eta_{1}\rangle\\
&=&\nonumber\langle\eta_{1}|\overline{\mu}(x)P_{acc}\overline{\mu}(x)^{\dagger}|\eta_{1}\rangle\\
&=&\nonumber\langle\psi_{0}^{(1)}|\overline{\mu}_{1}(x)P_{acc,1}\overline{\mu}_{1}(x)^{\dagger}|\psi_{0}^{(1)}\rangle\\
&=&P_{{\cal A}_{1}}(x)
\end{eqnarray}
and
\begin{eqnarray}
P_{\eta_{2}}(x)&=&\nonumber\|\langle\eta_{2}|\overline{\mu}(x)P_{acc}\|^{2}\\
&=&\nonumber\langle\eta_{2}|\overline{\mu}(x)P_{acc}P_{acc}^{\dagger}\overline{\mu}(x)^{\dagger}|\eta_{2}\rangle\\
&=&\nonumber\langle\eta_{2}|\overline{\mu}(x)P_{acc}\overline{\mu}(x)^{\dagger}|\eta_{2}\rangle\\
&=&\nonumber\langle\psi_{0}^{(2)}|\overline{\mu}_{2}(x)P_{acc,2}\overline{\mu}_{2}(x)^{\dagger}|\psi_{0}^{(2)}\rangle\\
&=&P_{{\cal A}_{2}}(x).
\end{eqnarray}
Therefore, $P_{{\cal A}_{1}}(x)=P_{{\cal A}_{2}}(x)$ holds if and
only if
\begin{equation}
P_{\eta_{1}}(x)=P_{\eta_{2}}(x) \label{Th1}
\end{equation}
for any string $x\in \Sigma^{*}$.

On the other hand, we have that
\begin{eqnarray}
P_{\eta_{1}}(x)&=&\nonumber\|\langle\eta_{1}|\overline{\mu}(x)P_{acc}\|^{2}\\
&=&\nonumber\sum_{p_{j}\in Q_{acc}}|\langle\eta_{1}|\overline{\mu}(x)|p_{j}\rangle|^{2}\\
&=&\nonumber\sum_{p_{j}\in Q_{acc}}\langle\eta_{1}|\overline{\mu}(x)|p_{j}\rangle(\langle\eta_{1}|\overline{\mu}(x)|p_{j}\rangle)^{*}\\
&=&\nonumber\sum_{p_{j}\in Q_{acc}}\langle\eta_{1}|(\langle\eta_{1}|)^{*}\overline{\mu}(x)\otimes (\overline{\mu}(x))^{*}|p_{j}\rangle(|p_{j}\rangle)^{*}\\
&=&\langle\eta_{1}|(\langle\eta_{1}|)^{*}\overline{\mu}(x)\otimes
(\overline{\mu}(x))^{*}\sum_{p_{j}\in
Q_{acc}}|p_{j}\rangle(|p_{j}\rangle)^{*}
\end{eqnarray}
and
\begin{eqnarray}
P_{\eta_{2}}(x)&=&\nonumber\|\langle\eta_{2}|\overline{\mu}(x)P_{acc}\|^{2}\\
&=&\nonumber\sum_{p_{j}\in Q_{acc}}|\langle\eta_{2}|\overline{\mu}(x)|p_{j}\rangle|^{2}\\
&=&\nonumber\sum_{p_{j}\in Q_{acc}}\langle\eta_{2}|\overline{\mu}(x)|p_{j}\rangle(\langle\eta_{2}|\overline{\mu}(x)|p_{j}\rangle)^{*}\\
&=&\nonumber\sum_{p_{j}\in Q_{acc}}\langle\eta_{2}|(\langle\eta_{2}|)^{*}\overline{\mu}(x)\otimes (\overline{\mu}(x))^{*}|p_{j}\rangle(|p_{j}\rangle)^{*}\\
&=&\langle\eta_{2}|(\langle\eta_{2}|)^{*}\overline{\mu}(x)\otimes
(\overline{\mu}(x))^{*}\sum_{p_{j}\in
Q_{acc}}|p_{j}\rangle(|p_{j}\rangle)^{*}.
\end{eqnarray}
Therefore, Eq. (\ref{Th1}) holds if and only if
\begin{eqnarray}
&&\nonumber\langle\eta_{1}|(\langle\eta_{1}|)^{*}\overline{\mu}(x)\otimes
(\overline{\mu}(x))^{*}\sum_{p_{j}\in
Q_{acc}}|p_{j}\rangle(|p_{j}\rangle)^{*}\\
&=&\langle\eta_{2}|(\langle\eta_{2}|)^{*}\overline{\mu}(x)\otimes
(\overline{\mu}(x))^{*}\sum_{p_{j}\in
Q_{acc}}|p_{j}\rangle(|p_{j}\rangle)^{*}
\end{eqnarray}
for any string $x\in \Sigma^{*}$.

Denote
\begin{equation}
D(x)= \overline{\mu}(x)\otimes\overline{\mu}(x)^{*}
\end{equation}
where $D(x)$ is an $(n_{1}+n_{2})^{2}\times (n_{1}+n_{2})^{2}$
complex square matrix. Then the equivalence between ${\cal A}_1$
and ${\cal A}_2$ depends on whether or not the following equation
holds for all string $x\in\Sigma^{*}$:
\begin{eqnarray}
&&\nonumber\langle\eta_{1}|(\langle\eta_{1}|)^{*}
D(x)\sum_{p_{j}\in
Q_{acc}}|p_{j}\rangle(|p_{j}\rangle)^{*}\\
&=&\langle\eta_{2}|(\langle\eta_{2}|)^{*} D(x)\sum_{p_{j}\in
Q_{acc}}|p_{j}\rangle(|p_{j}\rangle)^{*} \label{Eq}
\end{eqnarray}
Consider the linear space $\mathbb{M}_{n^{2}}$ consisting of all
$(n_{1}+n_{2})^{2}\times (n_{1}+n_{2})^{2}$ complex square
matrices. It is clear that the dimension of $\mathbb{M}_{n^{2}}$
equals $(n_{1}+n_{2})^{4}=n^{4}$.

By ${\cal D}^{(i)}$ we denote the subspace of $\mathbb{M}_{n^{2}}$
spanned by $\{D(x): x\in\Sigma^{*}, |x|\leq i\}$, where $|x|$
denotes the length of $x$. Clearly, we have
\begin{equation}
{\cal D}^{(0)}\subseteq {\cal D}^{(1)}\subseteq\cdots\subseteq
{\cal D}^{(i)}\subseteq {\cal D}^{(i+1)}\subseteq\cdots.
\end{equation}
Since the dimension of ${\cal D}^{(i)}$ is not more than
$(n_{1}+n_{2})^{4}$ for any $i\geq 1$, there exists $i_{0}$ such
that for any $N\geq i_{0}$, ${\cal D}^{(i_{0})}={\cal D}^{(N)}$.
In the rest of the proof, our purpose is to fix $i_{0}$.

 For the sake of convenience, we denote
$A_{i}=\mu_{1}(\Lambda^{k_{1}-i}\sigma^{i})$ for
$i=1,2,\ldots,k_{1}$, and
$B_{j}=\mu_{2}(\Lambda^{k_{2}-j}\sigma^{j})$ for
$j=1,2,\ldots,k_{2}$.  Set $k=\max(k_{1},k_{2})$. If $k_{1}\leq
k_{2}$, then we denote $A_{i}=A_{k_{1}}$ for
$i=k_{1}+1,k_{1}+2,\ldots,k$; if $k_{2}\leq k_{1}$, then we denote
$B_{j}=B_{k_{2}}$ for $j=k_{2}+1,k_{2}+2,\ldots,k$.

In addition, we denote $C_{i}=A_{i}\oplus B_{i}$ for
$i=1,2,\ldots,k$. Then $C_{i}$ is an $n=n_{1}+n_{2}$ order unitary
matrix for $i=1,2,\ldots,k$. According to the definition of
$\overline{\mu}(x)=\overline{\mu}_{1}(x)\oplus
\overline{\mu}_{2}(x)$, we know that $C_{1}C_{2}\cdots
C_{i}=\overline{\mu}(x)$ for $x\in\Sigma^{*}$ and $|x|=i\leq k$.
On the other hand, if $i\geq k$, then $C_{1}C_{2}\cdots
C_{k}^{i-k+1}=\overline{\mu}(x)$.

Thus, $D(x)=(C_{1}C_{2}\cdots
C_{i})\otimes(C_{1}^{*}C_{2}^{*}\cdots C_{i}^{*})$ for
$x\in\Sigma^{*}$ and $|x|=i\leq k$; and if $i\geq k$, then
$D(x)=(C_{1}C_{2}\cdots C_{k}^{i-k+1})\otimes
(C_{1}^{*}C_{2}^{*}\cdots (C_{k}^{i-k+1})^{*})$.

We set $E^{(i)}=span\{D(x):x\in\Sigma^{*},k\leq |x|\leq k+i\}$,
$i=0,1,2,\ldots$. Then, by means of Lemma \ref{Lm} it follows
that, there exists $i_{0}\leq n^{4}-1$, such that
\begin{equation}
E^{(i_{0})}=E^{(i_{0}+s)}\label{Th2}
\end{equation}
for any $s\geq 0$.

 Eq. (\ref{Th2}) implies that, for any
$x\in\Sigma^{*}$ with $|x|\geq k+i_{0}$, $D(x)$ can be linearly
represented by some matrices in $\{D(y): k\leq |y|\leq k+i_{0}\}$.
Therefore, if  Eq. (\ref{Eq}) holds for $|x|\leq n^{4}+k-1$, then
so does it for any $x\in\Sigma^{*}$.  We have proved this theorem.
\qed\\

\begin{Rm}
We analyze the complexity of computation in Theorem 11. As in
\cite{Tze92}, we assume that all the inputs consist of complex
numbers whose real and imaginary parts are rational numbers and
that each arithmetic operation on rational numbers can be done in
constant time. Still we denote $n=n_{1}+n_{2}$.  Note that in time
$O(in^{4})$ we check whether or not Eq. (\ref{Eq}) holds for $x\in
\Sigma^{*}$ with $|x|=i$. Because the length of $x$ to be checked
in Eq. (\ref{Eq}) is at most $n^{4}+k$, the time complexity for
checking whether the two multi-letter QFAs are equivalent is
$O(n^{3}(1+2+\ldots+(n^{4}+k))$, that is at most
$O(n^{12}+k^{2}n^{4}+kn^{8})$.

\end{Rm}

\section*{5. Concluding remarks}

In this paper, we have considered several issues concerning
multi-letter QFAs. Our technical contributions mainly contain the
following two aspects: (1) we have shown that $(k+1)$-letter QFAs
are strictly more powerful than $k$-letter QFAs, that is,
$(k+1)$-letter QFAs can accept some regular languages unacceptable
by any $k$-letter QFA, and some examples of regular languages
unacceptable by multi-letter QFAs have been provided. We have
known that multi-letter QFAs are strictly more powerful than
MO-1QFAs \cite{MC00}, but they are not comparable to MM-1QFAs
\cite{KW97,AF98} since the language $a^{*}b^{*}$ can be accepted
with bounded error by MM-1QFAs but cannot be accepted by
multi-letter QFAs, and the language $(a+b)^{*}b$ shows the
opposite direction. Moreover, $a^{*}b(a^{2})^{*}a$ cannot be
accepted  by MM-1QFAs and by multi-letter QFAs with bounded error.
(2) We have proved that a $k_{1}$-letter QFA ${\cal A}_1$ and
another $k_{2}$-letter QFA ${\cal A}_2$ for accepting unary
languages are equivalent if and only if they are
$(n_{1}+n_{2})^{4}+k-1$-equivalent, and the time complexity of
this computing method is $O(n^{12}+k^{2}n^{4}+kn^{8})$,  where
$n=n_{1}+n_{2}$, $n_{1}$ and $n_{2}$ are the numbers of states of
${\cal A}_{1}$ and ${\cal A}_{2}$, respectively, and
$k=\max(k_{1},k_{2})$.

The method  presented in the paper may be generalized to deal with
more general cases.  Another issue worthy of consideration is
concerning the state complexity of multi-letter QFAs compared with
the usual 1QFAs for accepting some languages (for example, unary
regular languages \cite{Yu98,RS97}). Also, the power of
measure-many multi-letter QFAs, as the relation between MM-1QFAs
and MO-1QFAs, is worth being clarified. Whether or not
measure-many multi-letter QFAs can recognize non-regular languages
may also be considered in the future.

\par

\section*{Acknowledgements}

The authors are very grateful to the two referees and Professor
Okhotin for their invaluable comments and suggestions that helped
greatly to improve the quality of the original manuscript. Also,
we thank our group of quantum computing of SYSU for finding some
mistakes in the original manuscript.

\end{document}